\documentstyle[aps,twocolumn]{revtex}
\begin{document}
\draft

\noindent
{\bf {\Large Comment on ``The Bright Side of Dark Matter''}}

\bigskip
In a significant recent paper A. Edery \cite{edery} undertakes a new
study of light deflection in generalizations of general relativity (GR). He
claims to prove that any metric-based gravitational theory that proposes to
explain the flat rotation curves of disk galaxies without postulating dark
matter halos must conflict with observations of gravitational lensing by
galaxies and clusters of galaxies because any such theory inevitably make a
negative contribution to light deflection.  Here we show that some of the
basic steps of Edery's argument are invalid, and no such general result
obtains.

Edery considers the metric of a spherical mass $M$ to be given by $ds^2
= B(M,r)dt^2 - A(M,r)dr^2 -r^2 d\Omega^2$, with which he calculates the
deflection  of a light ray approaching from infinity and then receding to
infinity.  The agreement with the solar-system tests of general relativity
requires that $AB\approx 1$ to high accuracy in the vicinity of the sun.
Unjustifiably, Edery extends this requirement to galactic scale, where in
conjunction with the requirement of  flat rotation curves, it leads to
suppressed light deflection. (It is easy to see that
without the constraint $AB=1$, flat rotation curves and enhanced light bending
are consistent.) But solar--system tests do not constrain the form of $A$ or
$B$ on galactic scale (except in the context of a specific gravitation theory).
Consider, for example, a modified-gravity theory with a mass scale, $M_0$,
below which it merges with GR [i.e. $A^{-1}(M,r)=B(M,r)=1-2GM/r$ for $M<M_0$].
If $M_0$ is between a solar mass and galactic masses, all solar-system results
agree with those of GR, while the form of $A$ or $B$ on galactic scale depends
on the exact theory. Similarly, if the departure from GR occurs only above a
certain length scale, intermediate between interstellar and galactic scale,
$AB=1$ in the solar--system, there is clearly no anomalous contribution to
light bending within the solar system, while again nothing can be generically
deduced about $A$ or $B$ on galactic scale.

In this connection we note another deficiency in Edery's
arguments: he calculates the deflection angle accumulated along the ray's path
all the way from infinity, and much of the undesirable negative contribution
comes from the asymptotic region. However, Edery's assumed form of the metric
is only valid out to limited radii: for galaxies only to a few megaparsecs
where the growth of the gravitational potential saturates as the galaxy's
field merges with the cosmological one; and near the sun only to a tenth of a
parsec where the mean field of the galaxy takes over. Alternative gravity
theories are generically nonlinear, so one cannot consider the contribution of
the sun separately from its galactic environment. Edery has also failed to
realize that in solar light--deflection experiments, only the {\it difference}
of deflection angles for two light paths, one grazing the sun, and one passing
about one earth--sun distance away, is actually measured. In such difference
the contribution from large distances, so crucial to his point about negative
light deflection, tends to cancel out.  Therefore, it is easy to devise
phenomenologically valid theories in which solar--system and galactic
predictions are unconnected. Edery supposes such connection as unavoidable
because he fails to realize that (i) the field of a totally isolated mass is
phenomenologically relevant only up to a limited distance, and (ii) the  sun
and galaxies are sufficiently different in mass, size, etc., so as to permit
theories that describe the two cases with totally different metric
coefficients.  This occurs, for example, in theories in which, in the spirit
of MOND\cite{milgrom}, departure from standard gravitation sets in only below
a certain acceleration scale (which is of order of the sun's acceleration at a
fraction of the interstellar distance).

If an alternative theory exhibits $AB\neq 1$ on galactic scale,
no contradiction between flat rotation curves and enhanced light--bending need
appear.  Both desired features coexist in Sanders' stratified
theory\cite{sanders}, which also predicts the PPN parameters for solar system
tests in the measured ranges. We also note that the dark-matter plus GR
standard doctrine eludes Edery's argument for suppressed deflection by having
$AB\neq 1$ around galaxies (since the matter density even outside the visible
galaxy does not vanish), but $AB\approx 1$ near the sun where no dark matter
is needed. But Edery does not tell us why we cannot have a modified--gravity
theory that gives for the metric of the visible matter in the whole universe
exactly what GR gives with dark matter ? Indeed, his statements are not really
about theories, but about metrics coming from unspecified equations. By
claiming that he only assumes a metric theory (one that obeys the equivalence
principle), he is driven to the conclusion that there is no metric for the
world that gives at the same time flat rotation curves, enhanced light bending
by galaxies, and consistency with the solar--system tests. This would seem to
exclude the metric calculated from GR with dark matter - a symptom of the
untenability of Edery's sweeping claim.

We acknowledge support of the Israel Science
Foundation to JDB and MM and of MINERVA to MM.
\medskip

\noindent
J. D. Bekenstein\par
Racah Institute of Physics\par
Givat Ram, Jerusalem 91904 Israel
\bigskip

\noindent
M. Milgrom\par
Weizmann Institute of Science\par
Rehovot, 76100 Israel
\bigskip

\noindent
R. H. Sanders\par
Kapteyn Astronomical Institute\par
Groningen, the Netherlands


\begin{thebibliography}{10}

\bibitem{edery} A. Edery, Phys. Rev. Letters {\bf 83}, 3990 (1999).
\bibitem{milgrom} M. Milgrom, Astrophys. Journ. {\bf 270}, 365, 384 (1983).
\bibitem{sanders} R. H. Sanders,  Astrophys. Journ. {\bf 480}, 492 (1997).

\end{thebibliography}
\end{document}